# High accuracy and error analysis of indoor visible light positioning algorithm based on image sensor

**Shihuan Chen[2,†], Weipeng Guan[1,†,*], Zequn Tan[2,†], Shangsheng Wen[3],**

**Manxi Liu[2], Jingmin Wang[2], and Jingyi Li[2]**

1. Department of Information Engineering, The Chinese University of Hong Kong, Shatin, New Territories, Hong Kong, China; weipengguan@cuhk.edu.hk (W.G.);
2. School of Automation Science and Engineering, South China University of Technology, Guangzhou 510640,China; cnchensh@outlook.com (S.C.); aumxliu@mail.scut.edu.cn (M.L.); 201730661329@mail.scut.edu.cn (Z.T.); 1902288474@qq.com (J.W.); aulijingyi@mail.scut.edu.cn (J.L.)
3. School of Materials Science and Engineering, South China University of Technology, Guangzhou 510640, China; shshwen@scut.edu.cn (S.W.)

* Correspondence: gwpscut@163.com (W.G.); weipengguan@cuhk.edu.hk (W.G.);

† Shihuan Chen, Weipeng Guan, Zequn Tan are co-first authors of this article.Their contributions to the paper are consistent, and their ranking is based solely on the first letter.

**Abstract:** In recent years, with the increasing demand for indoor positioning service, visible light indoor positioning based on image sensors has been widely studied. However, many researches only put forward the relevant localization algorithm and did not make a deep discussion on the principle of the visible light localization. In this paper, we make a deep discussion on the principle of the two-light positioning algorithm and the three-light positioning algorithm based on the image sensor, which includes how these positioning algorithms work and the errors analysis. Based on the discussion above, we propose two methods to improve the positioning accuracy, which is rotation method and dispersion circle method respectively. In our experiment, we have numerically and experimentally verified the two optimization methods and we obtained good positioning results. Especially, the positioning accuracy of the dual-lamp positioning algorithm based on dispersion circle optimization is up to 1.93cm, while the average positioning error is only 0.82cm, which is state-of-the-art of the same type of positioning system at present.

**1. Introduction**

In recent years, with the increase of indoor parking lot, shopping mall and other large facilities, the demand for indoor high-precision positioning is higher and higher. However, it is difficult for the satellite positioning system to receive satellite signals through the wall, which makes it hard to realize the indoor positioning based on the satellite positioning system. In order to overcome this problem, indoor positioning algorithms based on wireless LAN, ultra-wideband (UWB), bluetooth and radio frequency identification (RFID) are proposed. However, most of them cannot meet the standard of high positioning accuracy and it is worse that some of them are too expensive to meet the needs of large-scale applications because they rely on extra expensive equipment. With the requirement of high precision, a high precision LED indoor positioning system based on VLC technology is proposed. Compared with other positioning systems, visible light positioning (VLP) has its special advantages. First of all, the visible light positioning system has the advantages of simultaneous lighting and positioning. Second, unlike indoor location services based on radio frequency (RF), VLP systems can take the advantage of a large unregulated visible spectrum without the interference from radio frequency communications.

At present, indoor positioning methods based on VLC can be divided into two categories: VLC based on photodiode (PD) [23,24]and VLC based on image sensor. PD is not an ideal positioning device for mobile terminals because the positioning based on PD will cause large errors due to the angle measurement, the received signal strength measurement, the light intensity variation and so on, and the accuracy of PD largely depends on the direction of beam. On the contrary, using image sensor as a receiver is a better choice. Firstly, visible light positioning based on image sensor does not have the above-mentioned defects of PD. Secondly, commercial mobile phones are usually equipped with image sensors, which makes the visible positioning system based on image sensors do not have to pay for the extra receiver and therefore further reduces the application cost of the visible positioning system.

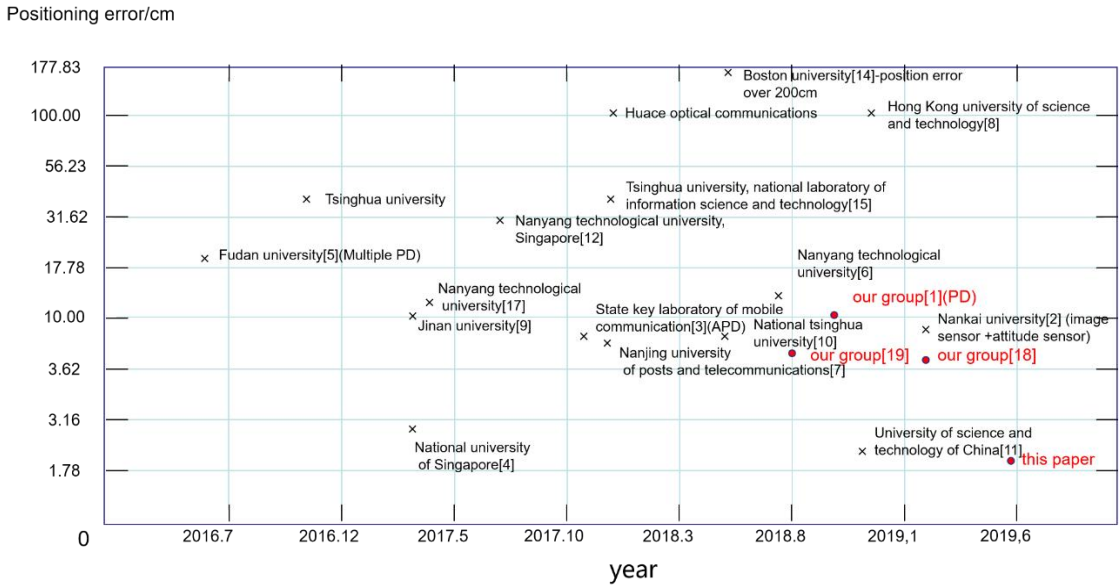

Fig. 1. the visible light indoor positioning accuracy of different institutions.

At present, many VLP methods have been proposed. We compared and analyzed the maximum positioning error of those visible light positioning papers [1-19] shown in figure 1, which were experimental results or except simulation. In [1], a high-precision indoor 3D positioning algorithm based on VLC and fingerprint identification is proposed, and the k-means and random forest algorithm are integrated. The maximum positioning error is 10 cm. In [2] the imaging principle of the camera under horizontal and oblique conditions is analyzed, and the positioning accuracy of 8cm is realized. In [25], a positioning system framework based on single LED lamp is introduced, which uses commercial mobile phones as receivers. The maximum positioning error of experiment is less than 7.39cm. In [3], the distance between the target and the reference LED is estimated by measuring the actual received signal strength (RSS). Then, relative distance and ternary number method are used for position estimation, and the positioning accuracy reached 6.6cm. In [4], FIR filter is used at the receiving terminal, and the predetermined normalized distance relation of received power is combined with the received power to calculate the distance. Two-dimensional positioning algorithm is used to determine the coordinates of the receiver, and the maximum positioning error is only 3cm. In [5,6], a reverse indoor positioning system based on VLC is proposed. Unlike other VLC positioning systems, the system installs multiple photodiodes on the ceiling to locate the person carrying the LED. The basic principle is to use the multi-photodiode to calculate the angle and the RSS to calculate the distance. This method can achieve the maximum positioning errors less than 20cm [5] and the average positioning accuracy of 7.4cm [6]. In [7], a positioning scheme based on the angle difference of arrival (ADOA) is proposed. The ADOA based positioning does not require the receiver to know its inclination angle. The positioning estimation is modeled as a three-variable optimization problem in the limited search area. This positioning scheme is able to achieve the positioning accuracy of 5.9cm. In [8], it is considered that there is a non-linear dependence of the received visible light intensity on user geometric position parameters, and LiFi positioning is a non-convex problem. By studying the inherent convex structure of the problem, an effective closed-loop updating equation for the user position and positioning angle is given, and the positioning accuracy is 100cm. In [9], a fine scintillation free line coding scheme and a lightweight image processing algorithm are designed to achieve the positioning accuracy of 10.7cm. In [10], coordinate data in the key format is transmitted through light-emitting diodes based light, and is captured by the smartphone camera. A novel perspective-n-point problem algorithm is used to estimate the position of the camera. This algorithm achieved a positioning accuracy of 6.22cm, but the experiment is only carried out at a height of 50cm. In [11], two positioning methods are proposed, one is with the known location of LED and the other is with the unknown location of LED. When the position of LED is unknown, the shift rotation model of receiver motion is established, and a new indoor positioning algorithm is proposed. The algorithm does not take the image center as the rotation center, but estimates the rotation center in the image, and the average location is only 1cm. In [12], a special single LED positioning method is proposed. The LED image is no longer processed as a point as in the existing work and it is processed as an image that uses geometric features to determine the direction and position of the receiver relative to the reference LED lamp. This method achieved a positioning accuracy of nearly 30cm. In [13], differential phase difference of arrival (DPDOA) algorithm which does not need local oscillators by using the differential phase difference method is proposed. In addition, kalman filter is used to reduce the change of range difference estimation and it achieves a maximum root-mean-square positioning error of 8cm. And [14] proposed a multi-unit lighting test bed for research and optimization of dense LiFi systems. [15] recovered the characteristics of optical wireless channels and used these characteristics to estimate the position of mobile terminals. Triangulation and nearest neighbor fingerprint algorithm are combined to reduce the computational complexity and the positioning accuracy is 40cm. In [16], an optical signal multiplexing technology using the Hadamard matrix for multiple-input multiple-output system is proposed. The actual experiment achieved 0.02m standard errors. [17] using singular value decomposition technology. In this technology, an approximate expression is derived to determine the position and direction of the receiver, and the positioning accuracy of 12cm is realized. [18] is our group's previous research on dual-light positioning algorithm based on image sensor, which achieved a positioning accuracy of 3.85cm. [19] is our group's research on three-light positioning algorithm, which achieved a positioning accuracy of 4.38cm.

In this paper, we deeply analyzed the previous work [18] on dual-light positioning system based on image sensor and [19] on three-light positioning system based on image sensor. In this process, the introduction of errors in the algorithm is emphatically analyzed. After eliminating some inevitable errors, it is concluded that the biggest source of errors in the algorithm is that the pixel coordinates of the image sensor are not accurate enough. In order to solve this problem, we propose two different optimization methods, rotation method and dispersion circle method. The experiment shows that the positioning accuracy of the optimized VLP system is greatly improved. Especially, the positioning accuracy of the dual-lamp positioning algorithm based on dispersion circle optimization reached 1.93cm, and the average positioning error is only 0.82cm. It can be said that with only an image sensor as the receiver and considering the measurement errors of lamps and cameras in positioning, the experimental results in this paper are state-of-the-art in the same type of positioning algorithm. The rest of this article is organized as follows. The second part introduces the system principle and the localization algorithm. The third part gives the experimental results and analysis of the system. And the final part is the conclusion.

## 2. System Principle
### 2.1. LED Positioning Algorithm

VLP system is mainly divided into two processes: LED-ID recognition and camera position calculation. We regard LED-ID detection and identification as a classification problem in the field of machine learning, and use fisher classifier and linear support vector machine to train the classifier offline and identify LED ID online. In the previous work [20-22], we had an deep discussion on the detailed process of LED-ID detection and identification. Due to the space constraints, this article only focuses on localization algorithms. For the readers that have interest in the LED-ID modulation and recognition, it is a great choice to refer to our previous reports [20-22].

The principle of visible indoor positioning algorithm based on image sensor is shown in figure 2. Generally speaking, the coordinates of LED in the world coordinate system are known, which are denoted as $(x_1, y_1, z_1)$, $(x_2, y_2, z_2)$ and $(x_3, y_3, z_3)$. Normally, the height of indoor ceiling is consistent and its value is equal to $z_1$ and $z_2$ and $z_3$. H is the vertical distance from the lens center of image sensor to the fixed point of LED, which is an unknown parameter. $f$ is the focal length of the image sensor and it is a fixed parameter of the camera. $(i_1, j_1)$, $(i_2, j_2)$ and $(i_3, j_3)$ is the coordinate of LED in the image coordinate system, which can be obtained by taking the coordinates of the LED on the pixel plane. The relationship between pixel coordinate system and image coordinate system is shown in figure 3.

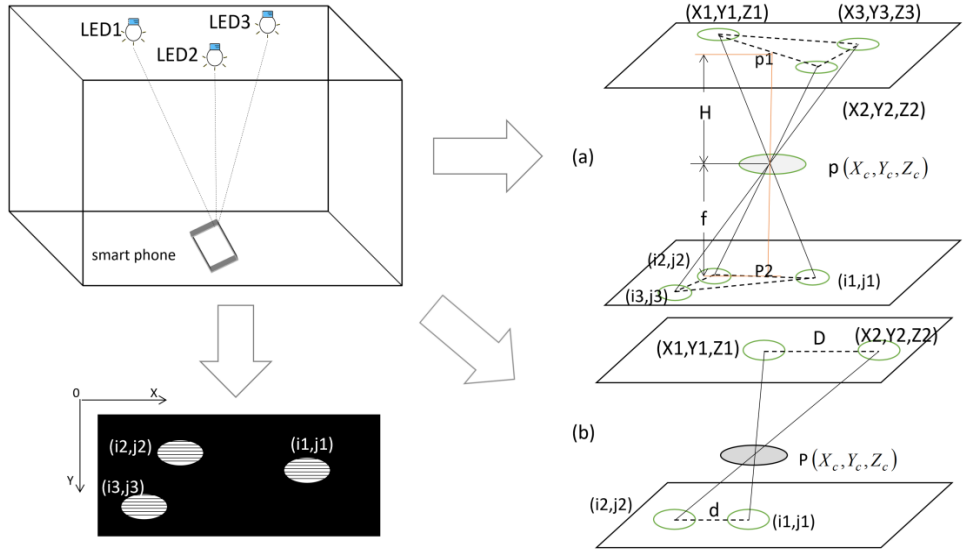

Fig. 2. Indoor positioning model based on LED lights.

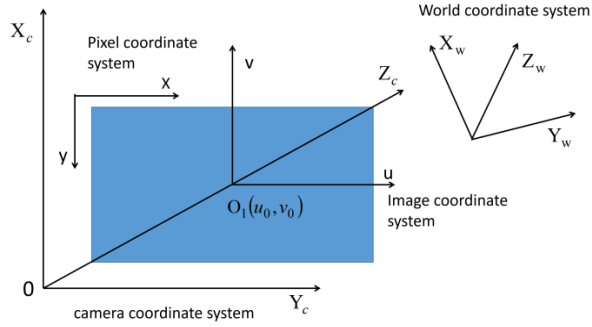

Fig. 3. The xy is the pixel coordinate system, vu is the image coordinate system, Xc Yc is the camera coordinate system, and Xw Yw Zw is the world coordinate system.

The origin point of image coordinate system is the intersection point between the optical axis of the camera and the imaging plane of the image sensor. It is also the center of the point in the imaging plane of the image sensor. The unit of the mentioned image coordinate system is mm(millimetre). And the unit of the pixel coordinate is pixel, described by the rows and lines of the pixels. Firstly, after obtaining the pixel coordinates of LED by camera and image processing, the coordinates of LED in the image coordinate system can be calculated according to the relationship between the pixel coordinate system and the image coordinate system:

$$i = (u - u_0)di \qquad (1)$$

$$j = (v - v_0)dj \qquad (2)$$

Where i,j are coordinates in the image coordinate system, u,v are coordinates in the pixel coordinate system, which can be obtained from the image through image process. di and dj represent unit transformation of two coordinate systems respectively, namely, 1pixel=dj mm. $(u_0, v_0)$ is the coordinates of the image sensor in the pixel coordinate system, which is in the center of the image. Therefore we can get the coordinates of LED in the image coordinate system. Then we can get the distance d of the LED in the image coordinate system and the distance D of the LED in the world coordinate system, which are shown in figure 2 (b)

$$d = \sqrt{(i_1 - i_2)^2 + (j_1 - j_2)^2} \qquad (3)$$

$$D = \sqrt{(x_1 - x_2)^2 + (y_1 - y_2)^2} \qquad (4)$$

According to the similar triangle principle, the vertical distance H between the lens center of image sensor and the fixed point of LED can be obtained. Then the z coordinate of image sensor can be obtained. $f$ is the focal length of the image sensor.

$$H = \frac{D}{d} f \tag{5}$$

$$z_c = z_1 - H = z_2 - H \tag{6}$$

Since the image coordinate system is established with the image sensor as the origin point, we can get the distance Li (i=1,2,3) between the LED and the image sensor in the image coordinate system.

$$L_i = \sqrt{i_i^2 + j_i^2} \tag{7}$$

According to the similar triangle principle, the horizontal distance Ri (i=1,2,3) between LED and image sensor in the world coordinate system can be obtained:

$$R_i = \frac{H}{f} L_i \tag{8}$$

Ri can also be described as:

$$\begin{cases}(x_c - x_1)^2 + (y_c - y_1)^2 = R_1 \\ (x_c - x_2)^2 + (y_c - y_2)^2 = R_2 \\ (x_c - x_3)^2 + (y_c - y_3)^2 = R_3\end{cases} \tag{9}$$

From equations (9) we can get these following two equations:

$$2(x_2 - x_1)x_c + 2(y_2 - y_1)y_c + (x_1^2 + y_1^2 - x_2^2 - y_2^2) = R_1 - R_2 \tag{10}$$

$$2(x_3 - x_1)x_c + 2(y_3 - y_1)y_c + (x_1^2 + y_1^2 - x_3^2 - y_3^2) = R_1 - R_3 \tag{11}$$

The equations above can get the coordinates of unknown nodes $x_c$ and $y_c$:

$$\begin{bmatrix} x_c \\ y_c \end{bmatrix} = \frac{1}{2}\begin{bmatrix} x_2 - x_1 & y_2 - y_1 \\ x_3 - x_1 & y_3 - y_1 \end{bmatrix}^{-1} \begin{bmatrix} R_1 - R_2 - (x_1^2 + y_1^2 - x_2^2 - y_2^2) \\ R_1 - R_3 - (x_1^2 + y_1^2 - x_3^2 - y_3^2) \end{bmatrix} \tag{12}$$

When there are three lights, the 3d coordinates of the camera $(x_c, y_c, z_c)$ can be obtained through formula (12) and (6). Nevertheless, there is no x3 and y3 when there are only two lights, hence the formula (12) cannot be solved. However, we can first assume that the image coordinate system and the world coordinate system are parallel and in the same direction. As shown in figure 2 (b), the positioning results can be obtained from the actual position and image position of the two LED lights:

$$\frac{x_c - \frac{x_1 + x_2}{2}}{H} = \frac{\frac{i_1 + i_2}{2}}{f} \tag{13}$$

$$\frac{y_c - \frac{y_1 + y_2}{2}}{H} = \frac{\frac{j_1 + j_2}{2}}{f} \tag{14}$$

However, the results of equations (13) and (14) are the world coordinates of the camera $(X_C, Y_C, Z_C) = (X_W, Y_W, Z_W)$ only when the two coordinate systems are parallel and in the same direction. But it is usual that there is no parallel relationship between the camera coordinate system and the world coordinate system. If the image coordinate system are not in line with the world coordinate system, the results obtained from equations (13) and (14) cannot be regarded as the world coordinates of the camera. As shown in figure 4, the rotation angle of image coordinate system and world coordinate system on the x and y planes needs to be taken into account to modify the result of formula (13),(14). Without the loss of generality, considering that the dual LED are parallel to the X-axis of the world coordinate system, the image coordinate of LED is relative to the image coordinate system, which can be shown in figure 4.

$$\theta = \operatorname{atan2}(j_1 - j_2, i_1 - i_2) \tag{15}$$

$$\begin{bmatrix} X_w \\ Y_w \\ Z_w \end{bmatrix} = \begin{bmatrix} \cos\theta & -\sin\theta & 0 \\ \sin\theta & \cos\theta & 0 \\ 0 & 0 & 1 \end{bmatrix} \begin{bmatrix} X_c \\ Y_c \\ Z_c \end{bmatrix} \tag{16}$$

From formula (16), the position of the camera based on the world coordinate system $(X_W, Y_W, Z_W)$ can be obtained. For more detailed description of our algorithm, please refer to our previous work[18,19].

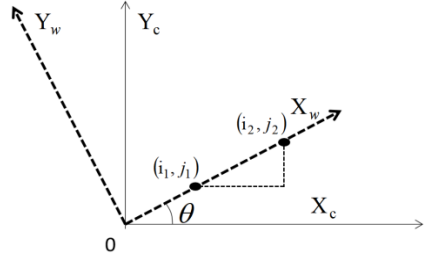

Fig. 4. The transformation model of camera coordinate system and the world coordinate system.

## 2.2. Algorithm analysis and optimization
### 2.2.1. Error analysis of the algorithm

In general, we believe that the VLP algorithm based on three lights will be better than the visible light positioning algorithm based on two lights. However, the previous experimental results [18,19] show that the average positioning error of the VLP algorithm based on double lamps is 1.99cm, while the average positioning error of the visible light positioning algorithm based on three lamps is 2.14cm. Therefore, the positioning accuracy of the visible light positioning algorithm based on two lamps is superior to the one based on three lamps, which is counterintuitive. Based on the fact above, we analyze the algorithm from the angle of error introduction to find why double lamp algorithm is better than three lamp. Firstly, the measurement error of LED position in the world coordinate system is eliminated. Because in the experiment, the actual position measurement of LED is accurate, even if error is introduced, it is extremely small, which will not be the main reason for the difference in positioning accuracy between the two algorithms. In the same way, the errors caused by the Pixel coordinates of LED are also can be ignored because the pixel coordinates of LED obtained by image processing program are accurate. Therefore, the main reason for the errors generation is the calculation process. In the algorithm, the coordinates of LED are based on the image coordinate system rather than the pixel coordinate system. And there is a big error in the process of transforming the coordinates of LED from the pixel coordinate to the image coordinate. The double lamp positioning only needs two LEDs, and the three lamp positioning algorithm needs three LEDs, which makes the error of the double lamp positioning algorithm less than that of the latter. This is the explanation of double lamp positioning algorithm is better than three lamp positioning algorithm. The following will explain why in the algorithm mentioned above the coordinates of LED have a large error in the process of converting from pixel coordinates to image coordinates. We can see that the camera's pixel coordinates (u0,v0) are used in the LED coordinate transformation equations (1) and (2). In our previous experiments, we assumed that the camera's pixel coordinates are at the center of the image.

However, we have verified that the camera's pixel coordinates are not at the center of the image. As shown in figure 5, O1 is the theoretical position of the image sensor. We spin the image sensor around the center of the lens and record the position of the LED lamp every 30 degrees or so. Obviously, the position of the LED in the image will be surrounded by a circle, and the center of the circle is the pixel coordinate of the image sensor. As shown in figure 5, the purple point is the initial position of the LED. Theoretically, we believe that the position of the LED lamp will rotate along the red track. But in fact, the rotation path of LED is a black circle. In other words, the position of image sensor is not the theoretical point O1, but the point O2 next to the point O1. The main reason for this is that the camera manufacturing process causes the led plane not to be completely parallel to the camera lens. We think this is the main reason for error generation.

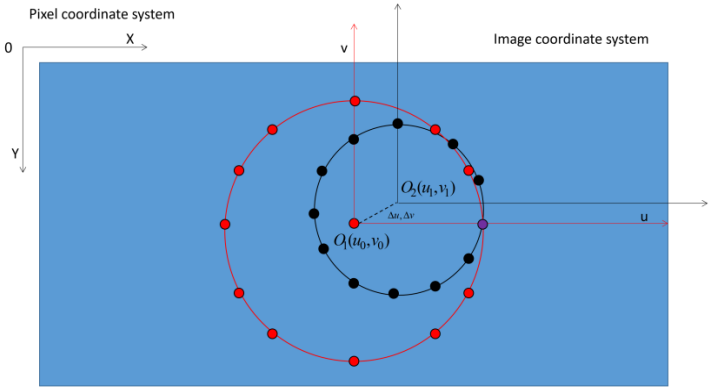

Fig. 5. Rotation method corrects image sensor position.

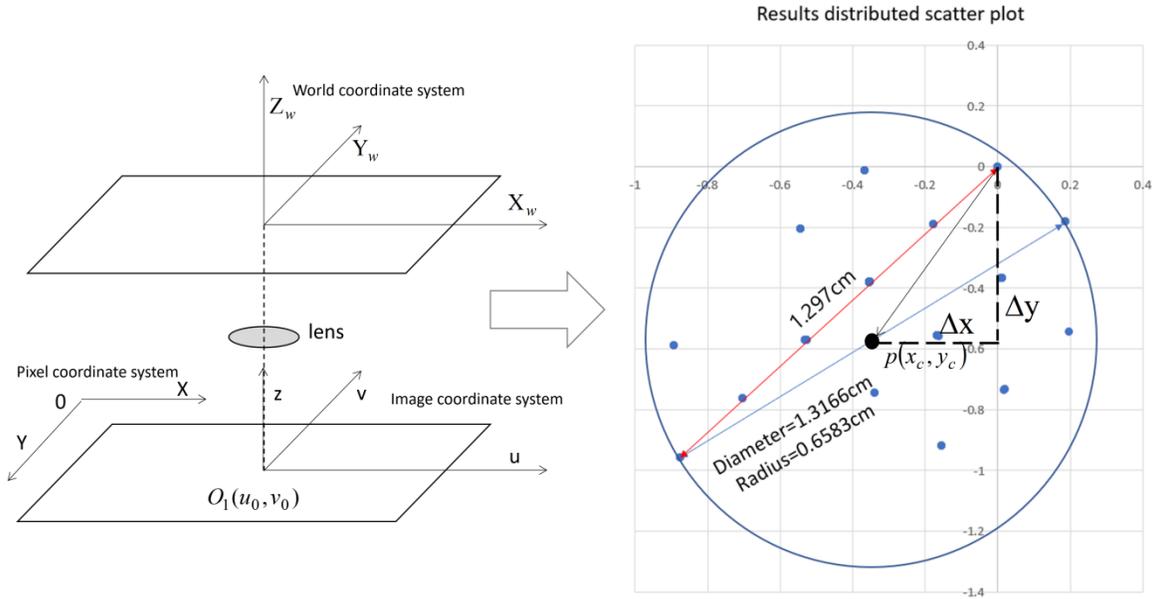

Fig. 6. Optimization schematic diagram based on dispersion circle.

**2.2.2. Algorithm to optimize**

Firstly, the rotation method mentioned in section 2.2.1 is used. After obtaining the coordinates of O2, we modified equations (1) and (2) into the following two equations. The remaining steps remain the same.

$$i = (u - u_1)di \qquad (17)$$

$$j = (v - v_1)dj \qquad (18)$$

In the experimental process of using rotation method, we find that the position of LED in the image does not exactly match a circle, which leads to the inaccurate position of the center of the circle. Even after correction, there is still error. Based on this situation, we use another method to correct the pixel coordinates of the image sensor. The optimization principle is based on the dispersion circle principle. The actual position of the image sensor remains unchanged, and the positioning results are different every time due to the existence of error. However, the positioning results are mostly distributed in a small area, and the diameter of the dispersion circle is the furthest distance between two points in all the results. We can assume that most of the positioning results of the system are in this circle. Ideally, when the center of the dispersion circle is corrected to coincide with the actual position coordinates, the euclidean distance between the data and the center of the dispersion circle is the actual positioning error, and most of the error do not exceed the dispersion radius. As long as the dispersion radius is small enough, the results can be very accurate when the camera is precisely calibrated. As shown in figure 6, we place the image sensor at the origin point of the world coordinate system and make enough measurements to find the smallest circle that can include all the results. The position of the center of the circle is the center of the dispersion circle. In practice, because the positioning results are generally about the symmetrical distribution of the center of the dispersion circle, the average value of all the results is taken as the position of the center of the dispersion circle for timeliness. So we can get the deviation between the center of the dispersion circle and the actual position, which is shown in the following equation(19) (20).

$$\Delta x = x_c - 0 = \bar{x} = \frac{1}{n}\sum_{i=1}^{n} x_i \tag{19}$$

$$\Delta y = y_c - 0 = \bar{y} = \frac{1}{n}\sum_{i=1}^{n} y_i \tag{20}$$

Based on the deviation, we use the following formula to correct the pixel coordinates of the image sensor:

$$u_1 = u_0 + \frac{\Delta x}{di} \tag{21}$$

$$v_1 = v_0 + \frac{\Delta y}{dj} \tag{22}$$

Finally, we use the traditional method of camera calibration to establish the correspondence between the known coordinate points on the calibration object and its image points through the calibration object whose size is known so that we can obtain the internal and external parameters of the camera model.

## 3. Experiment and Analysis

The optimization performance of the algorithm is verified by experiments. Our experimental platform is shown in figure 7. Five LED lights are installed on the top of the bracket. But only three LEDs are used. Each LED fixture is embedded with an 8-bit microcontroller unit to encode a unique identifier into a code word suitable not only for light transmission, but also for reducing flicker and dimming support. Using on-off key intensity modulation, LED drivers can convert code words into modulated digital signals, which drive LED to transmit light signals. The driving circuit board is shown in figure 8. In addition, we used CMOS industrial camera as positioning device, acer Aspire vn7-593g, Intel (R) Core (TM) i7-7700hq CPU @2.8ghz and Ubuntu 16.04 LTS as software platform to measure the performance of the algorithm. Table 1 provides all the key system parameters.

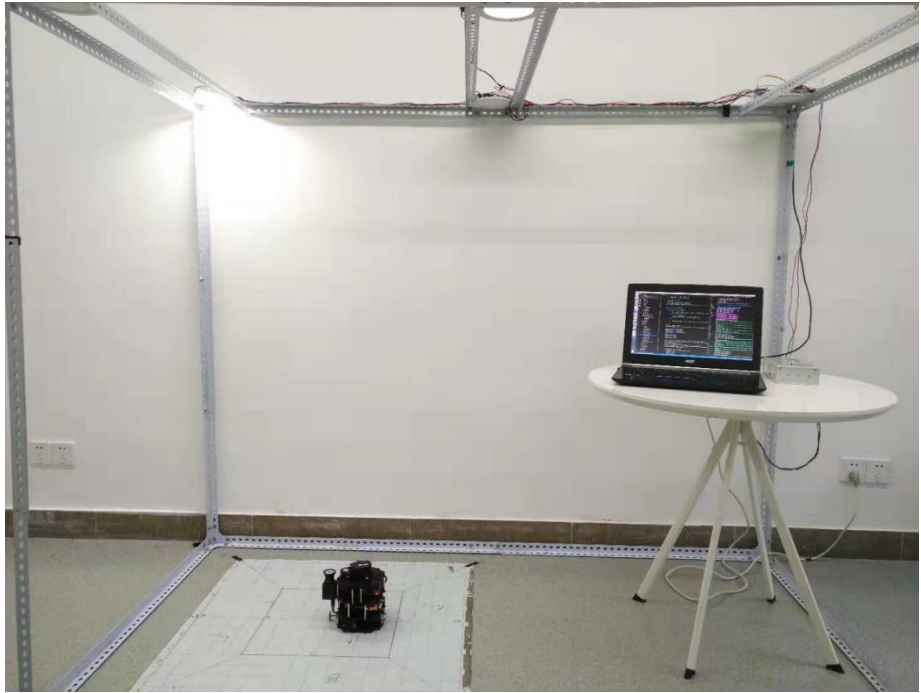

Figure 7. The experimental platform of VLP system based on positioning algorithm.

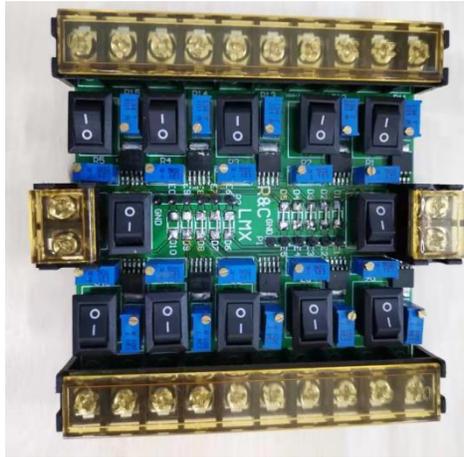

Figure 8. The circuit board of transmitting terminal.

**Table 1.** Parameters in this paper.

| Parameter | Value |
| --- | --- |
| Indoor space unit size(L × W × H) (m 3 ) | 2 × 1.1 × 1.6 |
| The focal length (mm) | 3 |
| Height of the camera (m) | 0 to 0.3 (resolution: 0.1) |
| Plan range of the camera (m) | 0.1 to 0.7 (resolution: 0.2) |
| Voltage of each LED (V) | 28.43 |
| Current of each LED (A) | 0.1 |
| The resolution of the camera | 800 × 600 |
| The exposure time of the camera (ms) | 0.05 |
| Computer parameter | Acer Aspire VN7-593G,Intel (R) Core (TM) i7-7700HQCPU@ 2.8GHz, Ubuntu 16.04 LTS |

As shown in figure.7, the performance of the optimization algorithm is verified by experiments, where the LED coordinates (cm) are (-46.5,-49.5,150), (-46,-42,150), (46,49,150) respectively. At height 0, there are 36 evenly distributed measuring points. Each position is tested 12 times. So after the test we can gain 432(12x36) results. The test steps of each point are as follows: First, we measure the actual position of the camera point by means of plumb and coordinate paper. Then, the image processing program segmented the LED image taken by the camera to obtain the position of the LED in the pixel coordinate system and used the linear classifier to identify LED-ID to obtain the position of the LED in the world coordinate system. Finally, the positioning results are obtained through the positioning algorithm.

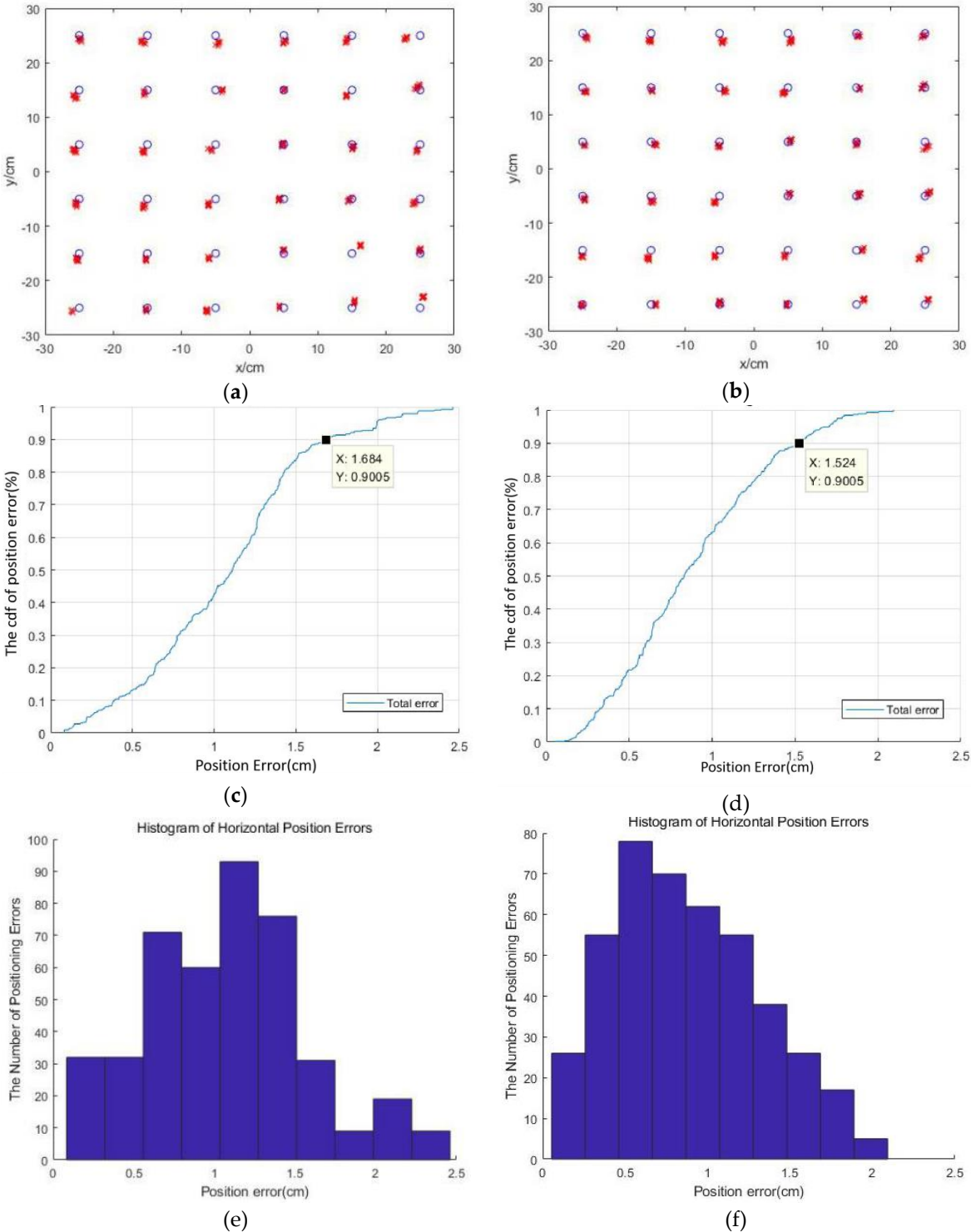

Figure 9. (a)The positioning results are based on the three - lamp rotating calibration method.(b)The result is based on the optimization of three-lamp spreading circle.(c)The CDF curves of positioning error in three lamp rotation optimization positioning system.(d)The CDF curves of positioning error in three-lamp distribution circle optimization positioning system.(e)Histogram of the positioning error in three lamp rotation optimization positioning system.(f)Histogram of the positioning error in three lamp rotation optimization positioning system.

The optimized positioning results based on the three-lamp rotation method are shown in Figure 9(a), and the optimized positioning results based on the three-lamp dispersion circle are shown in Figure 9(b). The blue circles represent the actual position of the camera, and the red crosses represent the positioning results. The distance between them is the positioning error of the VLP system. The experimental results show that the estimated position is well matched with the actual position and the high positioning accuracy can be achieved. Figure 9(c) and (d) are respectively the CDF curves after the two optimization, and Figure 9(e) and (f) are the positioning error histograms based on the two optimization methods. As can be seen from the CDF curves of Figure 9(c) and (d), the 90% positioning error of the three-lamp rotation method is less than 1.684cm, the average positioning error is 1.08cm, and the maximum positioning error is less than 2.46cm. In addition, based on the three-lamp dispersion circle optimization, the 90% positioning errors are less than 1.524cm, the average positioning error is 0.88cm, and the maximum positioning error is less than 2.10cm.

As for the positioning error of the optimized algorithm, some of them are caused by human factors. Firstly, when determining the actual position of the camera through the manually drawn network, there will be some errors between the predetermined position and the actual position of the camera. Secondly, when installing the LED lamp, the position of the LED cannot completely conform to our preset position, that is, the coordinate of the LED in the world coordinate system has some errors. Finally, the correction of camera pixel coordinates. Even if we have optimized the problem through rotation method or dispersion circle method, a small amount of errors will generate when the lamp track fitting is used to find the center of the circle. In the dispersion circle, since we use the average of the positioning results instead of the center of the circle, this process also introduces a small amount of error.

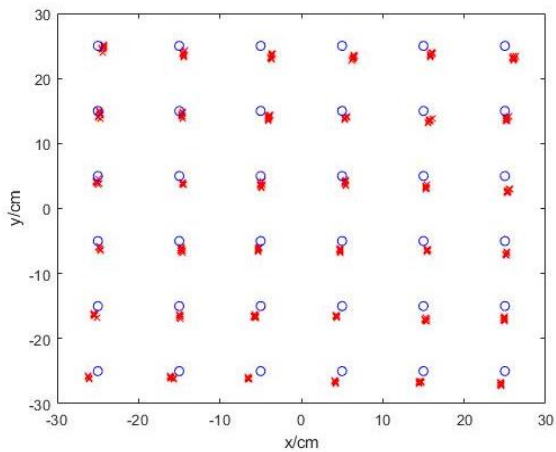
(a)

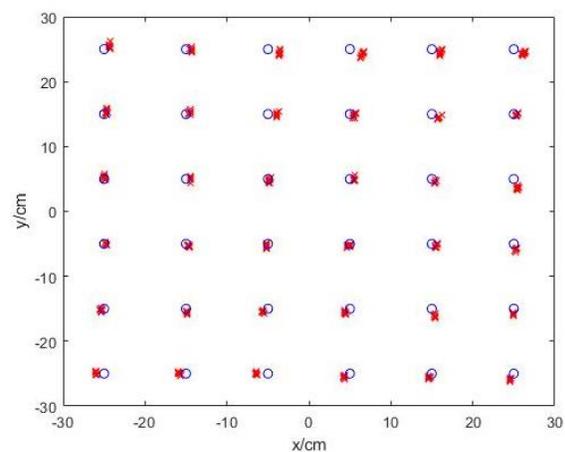
(b)

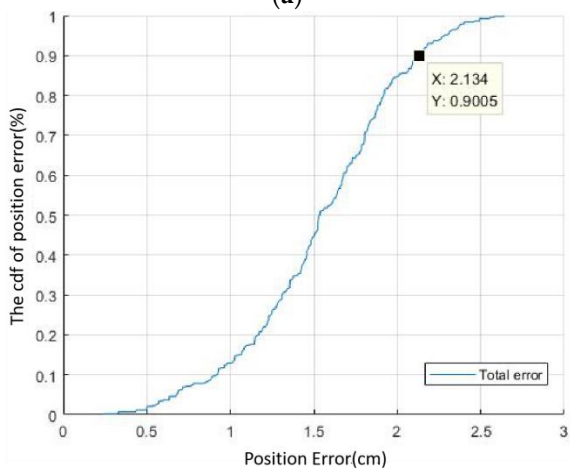
(c)

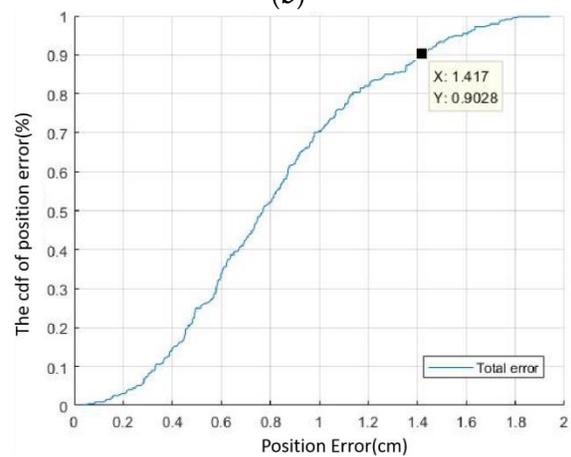
(d)

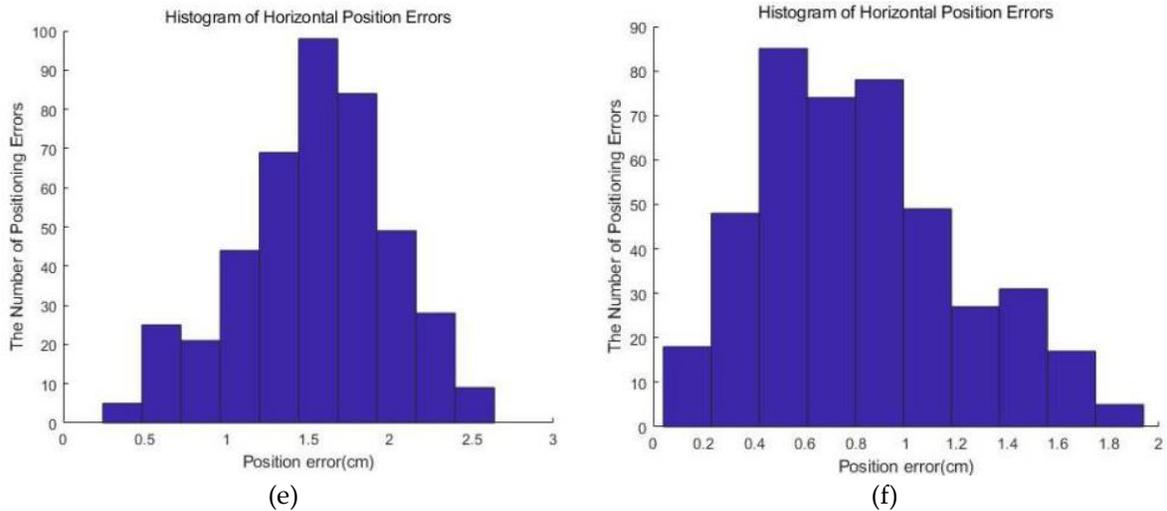

(e) (f)

Figure 10. (a)The positioning results are based on the two - lamp rotating calibration method.(b)The result is based on the optimization of two-lamp spreading circle.(c)The CDF curves of positioning error in two lamp rotation optimization positioning.(d) The CDF curves of positioning error in two-lamp distribution circle optimization positioning system.(e)Histogram of the positioning error in two lamp rotation optimization positioning system.(f)Histogram of the positioning error in two lamp rotation optimization positioning system.

Figure 10(a) is the optimized positioning result of double lamp rotation method and Figure 10(b) is the optimized positioning result of double lamp dispersion circle. Figure 10(c) and (d) are the CDF curves of the two optimization. Figure 10(e) and (f) are the positioning error histograms based on the two optimization methods. As can be seen from the CDF curves in Figure 10(c) and (d), 90% positioning errors of the optimized dual-light rotation method are less than 2.134cm, the average positioning error is 1.54cm, and the maximum positioning error is 2.64cm. After optimization of the double-lamp dispersion circle method, the 90% positioning error is less than 1.417cm, the average positioning error is 0.82cm, and the maximum positioning error is 1.93cm. As for the effects of the two optimization methods, we believe that the dispersion circle optimization method is superior to the rotation method. From the results of the optimization algorithm, it can be seen that the dispersion circle method is superior to the rotation method. First of all, in terms of positioning accuracy, the positioning error of the spread circle optimization is less than the error of the rotation method in both the two-light positioning algorithm and the three-light positioning algorithm. Secondly, it can be seen from (e) and (f) of Figure 9 and 10 that the peak value of the positioning error using the dispersion circle method is more concentrated in the area close to 0, while the distribution of the positioning error using the rotation method is less concentrated, and there are more distributions in the outer edge. As for the main source of error, the influence of human factors on the algorithm is explained in the optimization algorithm of three lights, which makes the world coordinates of LED lights and the real position of the camera all have certain errors. In fact, these errors can be reduced by more accurate measurement methods.

The number of LED used by the dual-light positioning algorithm is less than that used in the three-light positioning algorithm. As show in Figure 4, in order to make up for the shortage of LED, the dual-light positioning algorithm uses the rotation angle between the image coordinate system and the world coordinate system in the calculation process. According to the formual(15), the calculation of this angle is also dependent on the image coordinates of the LED lamp, so the correction of the image center in the dual lamp positioning not only corrects the image coordinates of the LED, but also corrects the rotation angle of the image coordinate system and the world coordinate system. As for other camera correction methods, their advantages are not obvious, hence they are not introduced in this paper.

Errors in measurement, lighting installation and camera placement are inevitable, but our algorithm ensures that the positioning accuracy can be achieved in the presence of these errors. With all experimental errors mentioned above taken into consideration, the rotation method and the dispersion circle method are used to achieve more accurate positioning. With the rotation method optimization algorithm and dispersion circle method, dual lights positioning reduced the average positioning error from 1.99 cm to 1.54 cm and 0.82 cm respectively. And with the rotation method optimization algorithm and dispersion circle method,three lights positioning reduced the average positioning error from 2.14 cm to 1.08 cm and 0.88 cm respectively. Among the four experimental results, the dispersion circle optimization algorithm based on dual lights has the smallest positioning error.The average positioning error of 0.82 cm and the maximum positioning error of 1.93 cm.

## 4. Conclusions

This paper presents a centimeter-level precise positioning system based on image sensor and visible light LED. In this paper, the principle of dual-light positioning algorithm and three-lamp positioning algorithm based on image sensor is deeply and respectively analyzed. And the error generation in the algorithm is discussed. It is known that the pixel coordinate of the image sensor in the algorithm is not in the center of the image due to the problem of camera manufacturing process. Based on this fact, two optimization methods are proposed. Experiments show that both rotating method and dispersion circle method can optimize the VLP system to a great extent. In dispersion circle optimization algorithm based on dual lights 90% positioning errors are less than 1.417cm, the average positioning error of 0.82cm, and the maximum positioning error of 1.93cm. Hence the dispersion circle optimization algorithm based on dual lights can be called the state-of-the-art of the same type of positioning system at present. This experiment is carried out when the camera plane is parallel to the led plane. How to realize high-precision positioning in the case of camera tilt remains to be further studied.

**Acknowledgments**

This research is funded by College Students Innovation and Entrepreneurial Training Program number S201910561205, National Undergraduate Innovative and Entrepreneurial Training Program number 201810561218, 201910561166, 201910561164 Special Funds for the Cultivation of Guangdong College Students' Scientific and Technological Innovation("Climbing Program" Special Funds) grant number pdjh2017b0040, pdjha0028, pdjh2019b0037 and Guangdong science and technology project grant number 2017B010114001. ,"The One Hundred-Step Ladder Climbing Program" grant number j2tw201902166, j2tw201902189.